\documentclass[a4paper,12pt,twoside]{article}
\usepackage{amsmath,amssymb,graphicx,cite,delarray}
\usepackage[a4paper]{geometry}
\usepackage[bf]{caption2}
\renewcommand{\captionlabeldelim}{.}

\begin{document}

\begin{center}\begin{Large}PHYSICAL NATURE OF fcc-bcc MARTENSITIC 
TRANSFORMATION IN IRON BASED ALLOYS\end{Large}\end{center}

\begin{center}
M.P. Kashchenko
\end{center}

\begin{center}
Physics Chair, Ural State Forest Engineering University,\\ 
Sybirskiy trakt, 37, 620100, Ekaterinburg, Russia
\end{center}

\sloppy

\begin{abstract} The summary of the models offered by the author 
revealing features of the physical mechanisms controlling 
processes of martensite crystal formation is resulted. 
The rapid growth of a cooling martensite crystal is considered 
as a self-organized process controlled by the quasi-longitudinal 
lattice displacement waves (DW). It is shown, that processes of 
the heterogeneous nucleation and wave growth have the genetic 
connection in case of spontaneous $\gamma-\alpha$ martensitic transformation. 
The exposition of strain martensite formation is considered in the 
context of a cryston model.\end{abstract}
\section*{1. Introduction}
One of the most characteristic features of $\gamma-\alpha$ martensitic 
transformations (MT) is their diffusionless transformation mechanism, 
by means of cooperative rearrangement of the face-centered cubic (fcc) 
high-temperature $\gamma$ -phase (austenite) into the body-centered cubic 
(bcc) or body-centered tetragonal (bct) low temperature $\alpha$ -phase (martensite). 
The disclosure of the dynamical mechanisms and principles underlying the process 
of martensitic transformations would enlighten their intrinsic characteristics and 
physical nature, and can thus be regarded as one of the fundamental problems of metal 
physics. The reconstructive phase transitions, to which the MT is related, demonstrate 
pronounced features of the first-order transitions, namely, considerable temperature 
hysteresis (between the direct and reverse transformations) and thermal and volume effects. 
As is known [1] the spontaneous (cooling-induced), stress- induced and strain-induced 
$\alpha$ - martensites are distinguished. Processes of martensite nucleation in all 
cases are heterogeneous.

Theoretical research on the $\gamma-\alpha$ MT is mainly characterized by the parallel 
development of the lattice-geometrical, thermodynamic and wave approaches. Now it is 
obvious that only the wave approach would have the full potential for a comprehensive 
description of the dynamical aspects of the transformation process in the cases of spontaneous 
and stress-induced martensites. In the stage of rapid martensite crystal growth there exists 
a boundary area between the phases being characterized by intensive electron currents within 
coexisting strongly pronounced temperature and - even more important - chemical potential 
gradients. An electronic drift current leads to an inverted occupation of those pairs of 
electronic states being localized in the proximity of the s-surfaces in quasi-momentum space. 
This surfaces are defined by the condition that the projection of electronic group velocity 
towards the orientations of $\nabla T$ or $\nabla \mu$ must vanish at all points of the s-surfaces. 
The number of pairs of inversely occupied electronic states of the 3d-bands of iron 
is a macroscopic quantity. The process of generation of atomic displacement waves  
is energized by stimulated emission of phonons during transitions of the 
non-equilibrium 3d-electrons between the inversely occupied states. 
The microscopic theory of generation of waves is in detail stated in [2] (see also [3,4]).

The constructive description of strain-induced martensite is achieved in frameworks of 
cryston model (definition of cryston it has been entered in [5]). 
In this case the crystons (the shear carriers of superdislocation type) are direct carriers 
of threshold strain. Thus there is clear enough fathoming of physical mechanisms for all 
alternatives of the $\gamma-\alpha$ martensitic transformation in iron-based alloys. 
The purpose of the report to have given the evident representation being accessible not 
only for narrow experts about the simplest models that will allow, in author opinion, 
to optimize the further researches of martensitic transformations.
 
\section*{2. Waves controlling the growth of martensite crystal}

The displacement waves controlling the process of martensitic crystal growth are of 
the longitudinal type (or quasi-longitudinal) with frequencies of $\nu \sim 10^{10}$ s$^{-1}$ 
(region of hypersound) and amplitudes ensuring the required level of lattice 
deformation of $\varepsilon \sim 10^{-3}$ needed for initiation of the 
$\gamma-\alpha$ -martensitic transformation. 
The mode of initial excitation of waves during the nucleation stage of the $\alpha$ -phase 
is a hard mode [6,7]. The certain combinations of displacement waves are important 
but not separate waves. Thus for instance, the stage of rapid growth of a spontaneous 
(and stress- induced) martensitic lamellae is correlated with the propagation of a 
pair of perpendicularly oriented waves, stimulating the process of flat lattice 
deformation of a combined tensile-compressive type.

\begin{figure}[htb]
\centering
\includegraphics[clip=true,width=.6\textwidth]{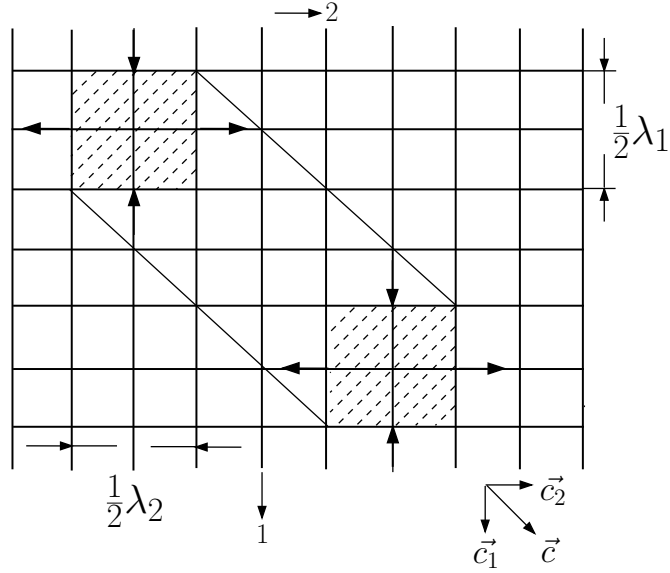}
\renewcommand{\captionlabeldelim}{.}
\caption{Fundamental growth pattern of a martensite lamella 
in the notion of two flat longitudinal waves propagating 
perpendicular to each other: $\mathbf{c}_{1}$, $\mathbf{c}_{2}$ - wave-velocities, 
$\lambda_{1}$, $\lambda_{2}$ - wavelengths.}
\label{plainwaves}
\end{figure}

It is easy to conceive from Fig.\ref{plainwaves} that the hatched area of wave-superposition, 
as well as the intersecting area of the wave-fronts, are simultaneously propagating 
at the velocity equivalent to the vector-sum (i.e. geometrical sum) of their individual 
velocities $\mathbf{c}_{1} \perp \mathbf{c}_{2}$, i.e. 
\begin{equation}
\label{eq1}
\mathbf{c} = \mathbf{c}_{1} + \mathbf{c}_{2},\quad c = \vert \mathbf{c}\vert = \sqrt{c_{1}^{2}+c_{2}^{2}}.			
\end{equation} 

As the value of $\vert \mathbf{c}\vert$, on the one hand, characterizes the frontal speed of growth of a 
martensite lamella, and on the other hand can exceed the longitudinal velocity of 
sound in direction of $\mathbf{c}$ in Fig.\ref{plainwaves}, it is possible in principle to explain this way 
the aligned supersonic growth of martensite crystals (being controlled by a pair 
of ordinary longitudinal waves), thus inherently representing an important kinetic 
particularity of the growth stage. From Fig.\ref{plainwaves} it is also obvious that the normal 
line to a habit plane of a growing crystal is set to the vector $\vec{N}$, being collinear 
with the normal vector of the plane defined by the vectors 
$\lbrack \mathbf{c}_{1}, \mathbf{c}_{2}\rbrack$ and $\mathbf{c}$, by 
the following vector-product
\begin{equation}
\label{eq2}
\mathbf{N}=\lbrack \mathbf{c}, \lbrack \mathbf{c}_{1}, \mathbf{c}_{2}\rbrack
\rbrack = \mathbf{c}_{1}\cdot c_{2}^{2}-\mathbf{c}_{2}\cdot c_{1}^{2}.     					
\end{equation}

If the concerning short waves are included into control waves system  the fine 
twinning may be described too [8,9].

\section*{3. Heterogeneous nucleation in elastic fields of dislocations}

The elastic field of dislocations disarranges the original lattice symmetry by 
selecting regions being most favorable for martensitic nucleation. Such a 
region features the shape of a perpendicular parallelepiped, its edges 
being oriented along the eigenvectors  $\vec{\xi}_{i}$ of the distortion tensor
$\hat{\varepsilon}$, its eigenvalues $\varepsilon_{i}$ satisfying 
the following conditions:

\begin{equation}
\label{eq3}
\varepsilon_{1} > 0,\quad \varepsilon_{2} < 0,\quad
\vert \varepsilon_{3}\vert \ll \vert \varepsilon_{1,2}\vert,      					
\end{equation}                                                         
thus ensuring the existence of slightly distorted  surfaces (SDS) with normals
\begin{equation}
\label{eq4}
(\mathbf{N}_{SDS})_{1,2}\:\Vert\:\vec{\xi}_{2}\:\mp\:
\vec{\xi}_{1}\sqrt{\frac{\varepsilon_{1}}{\vert \varepsilon_{2}\vert}},\quad
\vert \vec{\xi}_{1,2} \vert=1.
\end{equation}
                    
Obviously, from a point of view of minimization of elastic distortion energy, 
phase-coupling is supported by weakly distorted (with $\varepsilon_{3} = 0$ 
invariants) planes. 
Thus it would be reasonable to expect that the normal of the habit-plane of the 
martensite crystal should match with one $\mathbf{N}_{DS}$. In fact, among the $\mathbf{N}_{DS}$, there exist 
$60^{\circ}$-dislocations with lines $\langle 1\bar{1}0\rangle$ situated near $\langle 557 \rangle$, 
$\langle 225 \rangle$ and $30^{\circ}$-dislocations 
with lines $\langle 1\bar{2}1\rangle$ situated near  $\langle 259 \rangle$, $\langle 3\;10\;15 \rangle$, 
being evidence of certain 
differences among the NC of packet vs. explosive-martensite. Moreover, in the 
orientational relationship of the phase-lattice, there are included the slip-plane 
and the dislocation line, the latter one acting as a nucleation center, which 
suggests us to give preference to the Kurdjumov-Sachs- or Nishiyama-relationships, 
for various NC.

The question related to the orientation of macroscopic shear $\mathbf{S}$ will be resolved 
in conjunction with the choice of one of the two orientations of the normal
$\mathbf{N}_{SDS}$. 
For this aim, let us consider the notation of the distortional tensor in the elastic 
field, being represented as the sum of two diad products, and discriminate the part 
containing two addends:
\begin{equation}
\label{eq5}
\mathbf{S}_{1}\cdot \mathbf{N}_{1} + \mathbf{S}_{2}\cdot \mathbf{N}_{2},\quad
\vert \mathbf{N}_{1,2}\vert = 1.                                                               
\end{equation}
We recall  that the diad product $\mathbf{S}\cdot \mathbf{N}$ defines a 
deformation with an invariant plane, where $\mathbf{N}$ - normal of a plane 
and $\mathbf{S}$ - vector characterizing the deformation. Further 
considering that austenite is metastable at the beginning of the martensitic  
transformation at $M_{S}$ temperature, it is justified in the case of 
$\vert \mathbf{S}_{1} \vert > \vert \mathbf{S}_{2} \vert$,  
to surmise that the plane with the normal $\mathbf{N}_{1}$ is distinguished, and that the anticipated  
orientation of macroscopic shear is close to $\mathbf{S}_{1}$. And vice-versa, 
for $\vert \mathbf{S}_{2} \vert > \vert \mathbf{S}_{1} \vert$, the components 
$\mathbf{N}_{2}$ and $\mathbf{S}_{2}$ will be discriminated, respectively. 
The results of this approach are in good accordance with experimental 
results [10,11].

\section*{4. Synthesis of concepts of the heterogeneous nucleation and 
of the wave growth of martensite crystals}

\begin{figure}[htb]
\centering
\includegraphics[clip=true,width=.8\textwidth]{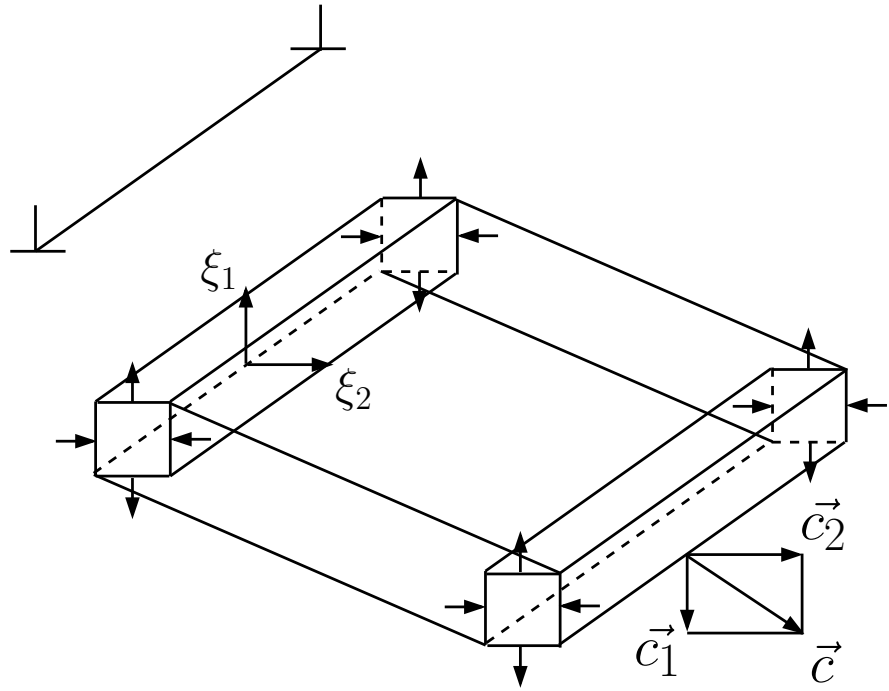}
\renewcommand{\captionlabeldelim}{.}
\caption{Nucleation and growth pattern of a martensite lamella.}
\label{fig2}
\end{figure}

With this approach, all macroscopic morphological characteristics of martensite 
attain a reasonable interpretation within the conceptual notion of nucleation at 
dislocations, where dislocations act as centers of forces disturbing the original 
lattice symmetry, their effect not being confined to the nuclear volume. These 
findings match in detail with the ideas of the wave theory of growth, presupposing 
that the transformation starts with the emergence of an excited state with the 
shape of a parallelepiped, built up of vectors $\vec{\xi}_{i}$, its pairs of edges oscillating 
in opposed phase, thereby  exciting controlling displacement waves orientated in 
the wave-normal $\mathbf{n}_{1,2}$ close to $\vec{\xi}_{1,2}$. In the most simple approximation 
of the equations 
\begin{equation}
\label{eq6}
\mathbf{n}_{1} =  \vec{\xi}_{1},\quad \mathbf{n}_{2} = \vec{\xi}_{2},
\end{equation}                                                                                                           
the requirement  of correspondence of $\mathbf{N}_{SDS}$ with the wave-habit (\eqref{eq2}) 
delivers the following condition:
\begin{equation}
\label{eq7}
\varkappa = \frac{c_{2}}{c_{1}} = \Big{(}\frac{\varepsilon_{1}}{\vert
\varepsilon_{2} \vert}\Big{)}^{1/2},
\end{equation}                                                                                  
which, if satisfied, ensures the possibility of a kinematic agreement 
of the wave description with the deformation description of the habit.
Fig.\ref{fig2} diagrammatically reflexes the process of nucleation in an 
elastic field of an edge dislocation and the growth controlling by the waves. 
As against Fig.\ref{plainwaves} the image three-dimensionally. 
The displacement waves exist in the shape of the wave bundles propagating 
in coordination with the spatially limited front of a  wave of relative 
volume deformation and making the function of a "pilot-waves", paving the way 
for the martensitic reaction in their wake.

Obviously, given the case that the ratio of tensile and compressive 
deformation in the wave-mode corresponds with $\varkappa^{2}$, then dynamic 
agreement will also be achieved. 
We further note that for the $\gamma - \alpha$ -transformation, which proceeds 
with increase of specific volume: $\varepsilon_{1} > \vert \varepsilon_{2} \vert$. 
Consequently, $c_{2} > c_{1}$, so that the tensile strain can be prescribed 
by the wave propagating with the smaller velocity $c_{1}$, whereas compressive 
strain can be prescribed by the wave propagating with the larger velocity $c_{2}$. 
(In the case of the $ \alpha-\gamma$ -transformation, the situation will just 
be inverted.)

\section*{5. Crystons controlling the growth of $\alpha$ strain 
induced martensite crystal}

\begin{figure}[htb]
\centering
\includegraphics[clip=true,width=.8\textwidth]{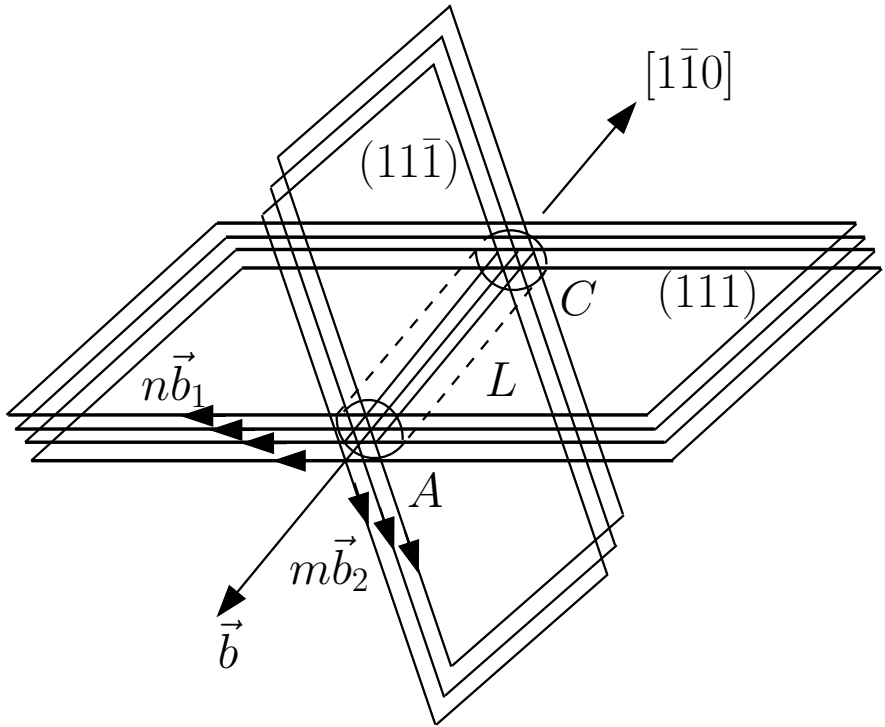}
\renewcommand{\captionlabeldelim}{.}
\caption{Typical for $\gamma$-phase generalized Frank-Read source of crystons 
with the total (superpositional) Burgers vector  $\vec{b}\:\Vert\:
n\vec{b}_{1}+m\vec{b}_{2}$.}
\label{fig3}
\end{figure}

The strain-induced martensite formation is considered as a consequence of carry 
of a threshold plastic deformation by crystons. The sources of crystons (carriers of shear 
of a superdislocation type) are caused by interaction of dislocations belonging to systems 
with intersected slip planes. Thus, the role of a pinned segment of an individual dislocation 
in the classic Frank-Read source is now played by a dislocation braid
(Fig.\ref{fig3}, region AC restricted by dashed lines), and the result of 
generation is a cryston (superdislocation) loop, which can be considered as a 
set of closed loops localized in the region bounded by a surface 
that is topologically similar to a torus.

As is shown in [12,13] the basic features of  the strain-induced martensite 
is described if to consider the cryston as the carrier of the threshold 
deformation having character of simple shear. 

\section*{6. Conclusion}

It is significant that concepts of the heterogeneous nucleation and the strain 
controlling the crystal growth (the last is localized in the frontal region 
of the growing crystal) are the universal for the exposition of martensite 
crystal formation. However the dynamic nature of controlling processes, as 
well as mechanisms of their energy support, essentially discriminate. 
For case of the cooling martensite this is the controlling wave process that 
is supported in the maser regime by nonequilibrium electrons due to the 
generation of energy in transforming phase. For case of the strain martensite 
this is the process of the cryston propagations that are supported 
in basic by energy of exterior stresses.

\section*{References}
\begin{enumerate}
\item[$\lbrack 1 \rbrack$] G. V. Kyrdjumov, L. M. Utevskii, R. I Entin, 
Transformations in steel and alloys, Nauka, Moscow, Russia, 1977, pp.41-98 (in Russian).
\item[$\lbrack 2 \rbrack$] M. P. Kashchenko, The wave model of martensite growth for the $\gamma-\alpha$ 
transformation in iron-based alloys, UIF Nauka, Ekaterinburg, Russia, 1993, pp. 139-168 (in Russian).
\item[$\lbrack 3 \rbrack$] V.P. Vereshchagin, M.P. Kashchenko, Principles of selection of pairs of 
electronic states which are  potentially active in phonon generation, Phys. Met. Metallog. 61,
$\mathcal{N}_{\:\:\bar{}\bar{}}^{\circ}\:2$ (1986) 23-30.
\item[$\lbrack 4 \rbrack$] M.P. Kashchenko, N. A. Skorikova, V. G. Chashchina, Conditions Required for Nonequilibrium Electrons
to Generate Elastic Waves in Metals with a Cubic Lattice. Phys.Met. Metallog. 99 (2005) 
$\mathcal{N}_{\:\:\bar{}\bar{}}^{\circ}\:5$.
\item[$\lbrack 5 \rbrack$] M. P. Kashchenko, V. V. Letuchev, L. A. Teplyakova, T. N. Yablonskaya, \\ A model of 
formation of macroshear bands and strain-induced martensite with (hhl) boundaries.Phys. Met. Metallog. 82 (1996) 329-336.
\item[$\lbrack 6 \rbrack$] M. P. Kashchenko, V. V. Letuchev, S. V. Konovalov, S. V. Neskoromnyi, Wave mechanism of 
growth and novel technique for initiation of the martensite nucleation.Phys. Met. Metallog. 76 (1993) 300-308.
\item[$\lbrack 7 \rbrack$] V.V. Letuchev, S.V. Konovalov, M.P. Kashchenko, Dynamical Lattice State at the 
Initial Stage of Martensitic Transformation and Possibilities of its Physical Realization.
Journal de Physique IV, Colloque C2, 5 (1995) 53-58.
\item[$\lbrack 8 \rbrack$] M. P. Kashchenko, V. G. Chashchina, Dinamicheskii mekhanizm dvoinikovaniia martensitnigo kristalla. 
Mekhanizmy deformatsii i razrusheniia perspektivnykh materialov. V.I. Betekhtin, S.P. Beliaev i dr. (Ed.) 
Sbornik trudov XXXV seminara "Aktual'nye problemy prochnosti", vol.1, PPI SPbGTU, Pskov, Russia, 1999, pp. 14-19 (in Russian).
\item[$\lbrack 9 \rbrack$] M. P. Kashchenko, V. G. Chashchina, Vliianie neodnorodnosti fronta 
upravliaushchego volnovogo protsessa na raspredelenie dvoinikov prevrashchenia v kristallakh 
martesita s gabitusami tipa (259) - (3 10 15).  V.G. Malinin (Ed.) Nauchnye trudy IV mezhdunarodnogo 
seminara "Aktual'nye problemy prochnosti v 2 tomakh", vol.1, NGU, Velikii Novgorod, Russia, 
2000, pp. 156-158 (in Russian).
\item[$\lbrack 10 \rbrack$] V.V. Letuchev, V.P.Vereshchagin, I.V. Alexina, M.P. Kashchenko, 
Conception of New Phase Dislocation-Based Nucleation at Reconstructive Martensitic Transformations.
Journal de Physique IV, Colloque C8. 5 (1995) 151-156.
\item[$\lbrack 11 \rbrack$] M. P. Kashchenko, V. V. Letuchev, S. V. Konovalov, T. N. Yablonskaya,
 Model of packet martensite formation. 
Phys. Met. Metallog. 76 (1997) 237-242.
\item[$\lbrack 12 \rbrack$] M. P.Kashchenko, V. G. Chashchina, A. G. Semenovih, Kristonnaia model' formirovania 
$\alpha^{\prime}$ martensita deformatsii v splavakh na osnove zheleza. Fizicheskaya. 
Mezomekhanika. 6 (2003) 95-122 (in Russian). 
\item[$\lbrack 13 \rbrack$] M.P. Kashchenko, A.G. Semenovih, V.G. Chashchina, Cryston model of strain induced martensite. 
J. Phys. IV France. 112 (2003) 147 - 150.
\end{enumerate} 
\bibliographystyle{unsrt}

\end{document}